\documentclass[aps,superscriptaddress,showpacs,preprintnumbers,amsmath,amssymb,floatfix]{revtex4-2}

\usepackage{graphicx,tabularx}% Include figure files
\usepackage{dcolumn}% Align table columns on decimal point
\usepackage{bm}% bold math
\usepackage{color}
\usepackage{color,soul}
\usepackage{epstopdf}
\usepackage{upgreek}
\usepackage{color}
\usepackage{xcolor}

\renewcommand{\figurename}{Fig.}

\begin{document}

%\preprint{APS/123-QED}

\title{High-Field and Light-Dependent Magnetization Provides Evidence for Coexisting Granular Superconductivity and Superparamagnetism in Nitrogen-Doped Diamond}

\author{E. Osmic}
\email{e.osmic@hzdr.de}
\affiliation{Hochfeld-Magnetlabor Dresden (HLD-EMFL), Helmholtz-Zentrum Dresden-Rossendorf, 01328 Dresden, Germany}

\author{J. Barzola-Quiquia} 
\email{j.barzola@gmx.de}
\affiliation{Felix-Bloch Institute for Solid-state Physics, University of Leipzig, 04103 Leipzig, Germany}
\affiliation{FHR Anlagenbau GmbH, Am H{\"u}gel 2, 01458 Ottendorf-Okrilla, Germany}

%\author{W. B{\"o}hlmann}
%\affiliation{Division of Superconductivity and Magnetism, Felix-Bloch Institute for Solid-state Physics, University of Leipzig, 04103 Leipzig, Germany}

\author{S. Winnerl}
\affiliation{Institute of Ion Beam Physics and Materials Research, Helmholtz-Zentrum Dresden-Rossendorf, 01328 Dresden, Germany}

\author{T. Herrmannsd\"{o}rfer}
\affiliation{Hochfeld-Magnetlabor Dresden (HLD-EMFL), Helmholtz-Zentrum Dresden-Rossendorf, 01328 Dresden, Germany}

\author{M. Uhlarz}
\affiliation{Hochfeld-Magnetlabor Dresden (HLD-EMFL), Helmholtz-Zentrum Dresden-Rossendorf, 01328 Dresden, Germany}

\author{J. Meijer}
\affiliation{Applied Quantum Systems Department, Felix-Bloch Institute for Solid-state Physics, University of Leipzig, 04103 Leipzig, Germany}

\date{\today}

\begin{abstract}
\section*{Abstract}
We investigated the field-, temperature-, and time-dependent magnetic moment of a nitrogen-doped type Ib diamond in darkness and under monochromatic light with a low laser intensity of about 160\,µWcm$^{-2}$. In darkness, below approximately 25\,K, the magnetic moment exhibits unconventional hysteresis loops that can be explained by the coexistence of superparamagnetism and superconductivity in certain regions of the sample. Under illumination, this hysteresis largely disappears, leaving mainly the superparamagnetic contribution. Temperature-dependent measurements reveal an irreversibility below 25\,K in darkness, which vanishes under illumination. Similarly, at 0.5\,T and below 25\,K, the time-dependent magnetization shows flux-creep-like behavior that disappears when the sample is illuminated. All light-induced changes are fully reversible: the magnetic properties measured in darkness are restored after the sample is heated above roughly 50\,K. 
Removing the superparamagnetic contribution from magnetization measurements up to $\pm 14$ T reveals clear characteristics of a granular superconductor—namely, a well-defined penetration field $H_p \approx 2.6$ T and a critical field $H_{c2} \approx 10.6$ T.
Based on theoretical considerations, we attribute the observed phenomena to the combined effects of internal stress and local variations in nitrogen concentration within localized regions that can be regarded as grains. This interpretation is supported by Raman and birefringence spectroscopy measurements on the same sample.
\end{abstract}

%\pacs{XX75.30.-m, 75.50.Pp,81.05.ug} %\verb+\pacs{#1}+
\maketitle

\section{Introduction}
\label{introduction}
Diamond, a natural allotrope of carbon, is transparent and insulating, a good thermal conductor at room temperature, and the hardest naturally occurring material on Earth. In the last years the interest in diamonds has immensely increased because of the promising
applications of the different color centers~\cite{WRACHTRUP-06}. Until today, the most prominent center is the negatively-charged nitrogen-vacancy (NV$^-$) center. 
The NV$^-$ center is a point defect in diamond consisting of a substitutional nitrogen with a vacancy in its nearest neighbor 
lattice site~\cite{DOHERTY-13}. NV$^-$ centers can be used as highly sensitive  magnetometers
with nanoscale resolution~\cite{TAY-08} and they offer a possibility of building a quantum computer working at room temperature~\cite{WEBER-10,PEZZAGNA-21}.
During the growth process of synthetic diamond  crystals  under ultrahigh-pressure and high-temperature (HPHT) conditions, 
nitrogen is easily incorporated into the diamond lattice
between $10...500$~ppm ($0.001\,\%...0.05\,\% $) and may act as a donor. 
Superconductivity in diamond was reached by heavy doping with Boron ($\approx 3\%$) as reported for the first time by Ekimov\,\textit{et al.}\,\cite{EKIMOV-04}, and it explained that the superconductivity is due to the coupling of the holes to the optical zone-center phonons, a mechanism similar to the one causing high-temperature superconductivity in MgB$_2$, but in three dimensions\,\cite{BOERI-04}.

Many theoretical works have studied the role of dopants in the superconductivity of diamond\,\cite{BLASE-04,LEE-04}, with the prediction that electron $n$-doped diamond can induce superconductivity~\cite{MA-05}. The theoretical calculation results indicated that nitrogen doping can turn diamond into a superconductor, suggesting that the donor impurity produces a stronger electron-phonon ($\lambda_{e-p}$) coupling, leading to higher critical superconducting temperatures $T_c$\,\cite{MA-05,COSTA-21}. It is worth emphasizing that according to the theoretical results, when diamond is doped with boron,
the $\lambda_{e-p}$ coupling rises essentially from the optical phonon modes. For nitrogen-doped diamond on the other hand, an intensifying acoustic contribution was observed as the nitrogen content was increased, leading to large values for
both the $\lambda_{e-p}$ coupling and $T_c$. This leads to the conclusion that between nitrogen and boron, the first may be a better impurity to turn diamond into a superconductor\,\cite{COSTA-21,MA-05}.
At the beginning of the study of superconductivity and their changes, e.g. influenced by the addition of magnetic and non magnetic dopands, high-frequency electromagnetic fields were also used \cite{DAYEM-67,ANDERSON-64}. 
Several processes can be induced by the influence of light on a superconductor. For example, the destruction of the superconductivity in Pb films  by laser light irradiation\,\cite{TESTARDI-71} was observed, significant changes on the magnetic moment and its temperature dependence of the ceramic superconductor LaBaCuO by visible light~\cite{LAIHO-90} and even the light induced superconductivity in the high temperature superconductor cuprates (e.g. YBCO and LaEuSrCuO \cite{FAUSTI-11,KAISER-17}) and in potassium doped fulleride K$_3$C$_{60}$~\cite{MITRANO-16,CANTALUPPI-18}. The pinning and motion of magnetic flux in superconductors are also influenced by light~\cite{TESTARDI-71,OWEN-72} as reported in YBCO\,\cite{LANDAU-95}, where weak visible light has extremely influenced the flux creep rate at low temperatures.

In this study, we have performed measurements on the magnetic properties of a nitrogen doped diamond crystal, in the dark and under the influence of illumination with monochromatic light. The measurements were performed by varying the magnetic field, temperature, and including time-dependent measurements of the magnetization. We used light to clarify the origin of the unusual magnetic properties observed below 25\,K in nitrogen-doped diamond\,\cite{JBQ_ND-2019,BARZOLA-22}.
 These results were obtained in the dark and interpreted as consequence  of the coexistence of superconductivity and superparamagnetism formed at small and dispersed regions inside the sample. In this work,  the laser wavelengths were chosen in order to investigate if some of the very well known nitrogen based
color centers in diamond are  involved with the unusual magnetic properties. 
We believe that our results support the previous interpretation, i.e. the coexistence of superconductivity and superparamagnetism, 
 where the illumination of the sample has strongly modified the magnetic properties of regions in the sample responsible for the hysteresis, i.e. the superconducting regions. Our results are reversible, i.e. after sample is heated, for example at temperatures above $50$\,K, the unusual magnetic properties observed in the dark are recovered.
The influence of light on the magnetic properties of our investigated sample is similar
to that observed in well known superconducting materials, supporting our interpretation 
of the coexistence of superparamagnetism and superconductivity in some regions of
the N-doped diamond sample. Additionally, to support our interpretation, we have done optical birefringence to show the existence of strained regions as consequence of remnant pressure and in addition, Raman spectroscopy, which allowed us to estimate the amount of nitrogen present in our sample as well as to confirm the presence of remnant pressure. Lastly, we measured the magnetization of a similar type Ib sample up to $\pm 14$\,T and
estimated a second critical field of $\approx 10$\,T, a value very similar to those reported for
boron-doped superconducting diamonds \cite{TAKANO-07,OKAZAKI-15}.

\section{Experimental}

\begin{figure}
\centering
\includegraphics[width=0.75\columnwidth]{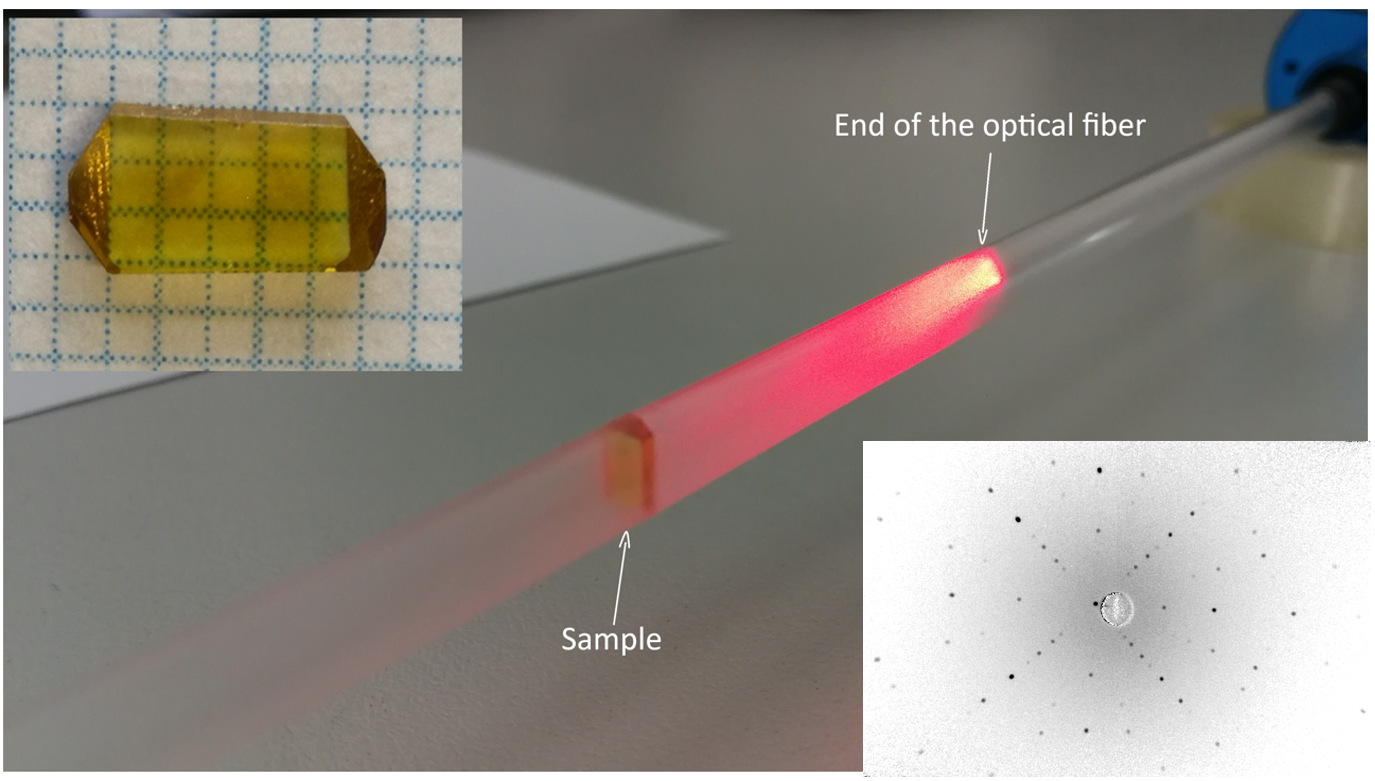}
\caption{\label{fig:fig1} SQUID sample holder showing the optical fiber and the diamond mounted inside the straw. The sample is illuminated with light of wavelength $\lambda_{L2} = 633$\,nm. The top-left image shows the studied diamond placed on graph paper, while the bottom-right image displays the X-ray Laue pattern measured along the same direction as the laser illumination.}
\end{figure}

We have investigated a nitrogen-doped type Ib diamond single crystal as received from the ELEMENTSIX company (Product number  145-500-0560). The  synthetic diamond crystal was  produced under ultrahigh-pressure and high-temperature (HPHT) conditions and the amount of dispersed nitrogen is in the order of hundreds of ppm,  according to the company. To clarify, the magnetization data presented in this work are shown exactly as measured by the SQUID magnetometer—that is, as raw data.
 An image of the diamond is shown in~\figurename\ref{fig:fig1}. Before the SQUID measurements, we removed surface contaminants by immersing the sample in a mixture of 30\,ml concentrated sulfuric acid (H$_2$SO$_4$), 10\,ml concentrated nitric acid (HNO$_3$), and 
10\,ml of 60\,\% perchloric acid (HClO$_4$). The solution was heated to about 130\,$^\circ$C under reflux for four hours. After cooling, the acids were decanted and the diamond was rinsed thoroughly with distilled water.  

Magnetic moment measurements were carried out using a Quantum Design SQUID magnetometer equipped with an optical fiber for \textit{in situ} illumination. Data were collected between 5\,K and 125\,K in magnetic fields up to 7~T. Monochromatic illumination was provided by diode lasers at 403, 633, and 840\,nm with powers of approximately 30, 26, and 25\,$\upmu$W, respectively. The laser power at the sample position was monitored using a THORLABS PM100D power meter with an S121C silicon sensor.
To determine whether the material exhibits type-II superconductivity, we searched for the presence of a second critical magnetic field. For this purpose, we used a diamond sample similar to the one investigated under illumination and measured it in a Quantum Design PPMS cryostat equipped with a superconducting magnet providing fields up to $14$\,T. A different sample was used ($1.5\times1.5\times1.5 ~\rm mm^3$) here due to the size constraints of the cryostat. X-ray Laue diffraction was carried out using an Inel EQUINOX 3000 diffractometer, with the X-ray beam aligned along the same direction as the illumination used during the magnetization experiments. The corresponding results are presented in the Supplementary Information (SI). We conducted room-temperature confocal Raman spectroscopy using a LabRAM HR Evolution spectrometer with a 488~nm excitation wavelength. Birefringence imaging was performed with a commercial Olympus BX51-P microscope.

%was used to verify the presence of color centers related to the nitrogen as
%nitrogen vacancies (NVs) and/or other color centers in our samples, the used laser of wave length $\lambda = 488$\,nm. 

\section{Magnetic properties measurements in the dark and under illumination }
\label{results}

\subsection{Field and temperature dependence}

We measured the field-dependent magnetic moment, $m(H)$, with the large sample surface oriented perpendicular to the applied magnetic field. Measurements with the field applied along other directions revealed no detectable anisotropy within the experimental resolution. 

\begin{figure}
\centering
\includegraphics[width=0.75\columnwidth]{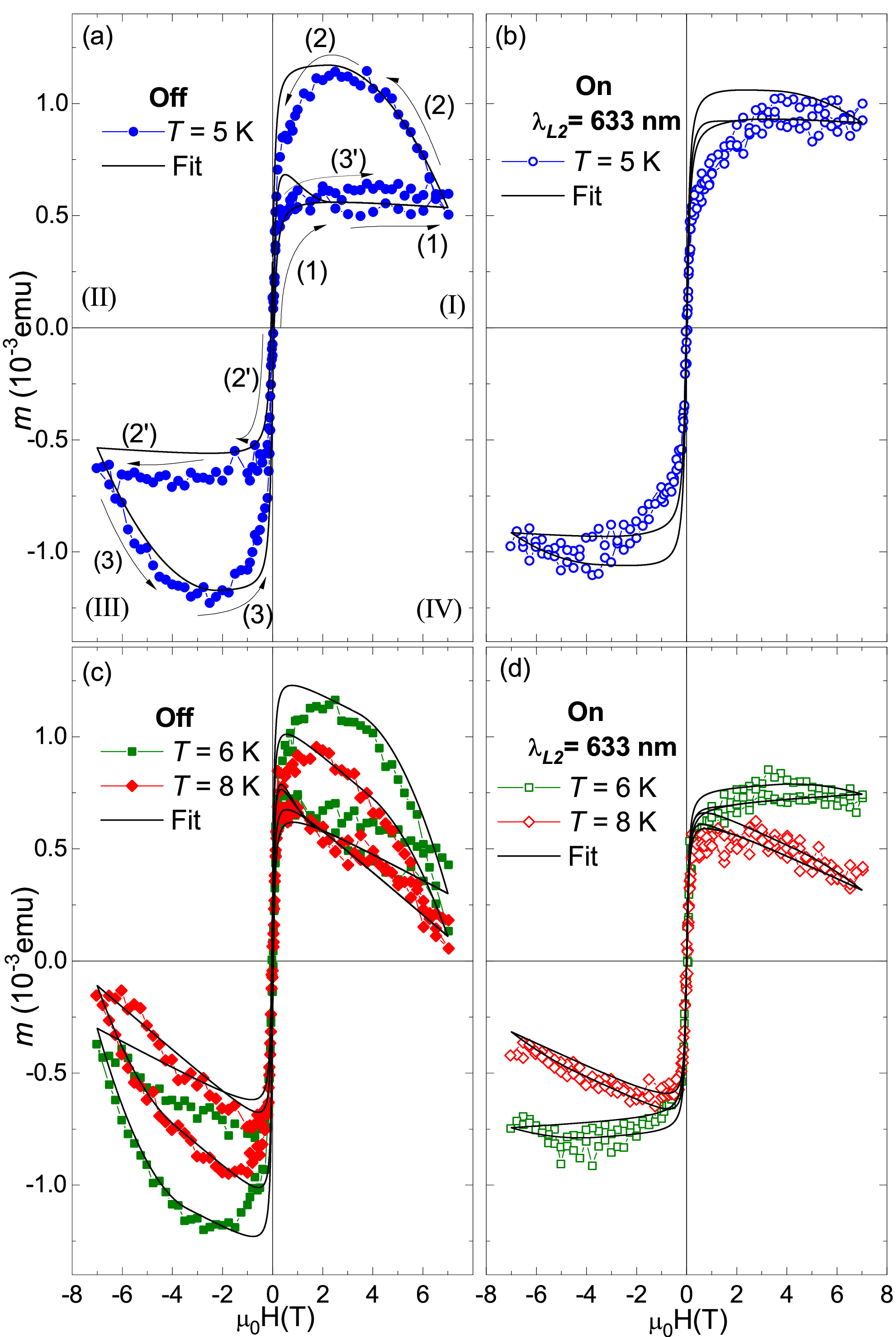}
\caption{\label{fig2} Magnetic moment $m$ as a function of the applied field measured at different constant temperatures. (a,c) Results measured in the darkness. (b,d) Results measured under illumination of light with $\lambda_{L2}= 633$~nm. Solid lines are fits done using equation~Eq.~\ref{eq:mtot}.}
\end{figure}

Hysteresis loops at constant temperatures were recorded by sweeping the field from $0\,\mathrm{T} \rightarrow +7\,\mathrm{T}$ (Step~1), then from $+7\,\mathrm{T} \rightarrow -7\,\mathrm{T}$ (Step~2), and finally back to $+7\,\mathrm{T}$ (Step~3), as indicated by the arrows in \figurename~\ref{fig2}(a). After each full cycle, the field was reduced to zero, and the sample was warmed to $125\,\mathrm{K}$ before repeating the measurement at the next temperature. 

For measurements under illumination, the sample was first heated to $125\,\mathrm{K}$, the laser was switched on, and after $15$~minutes of stabilization, the sample was cooled to the desired temperature. Results obtained in darkness and under illumination with light of $\lambda_{L2}=633\,\mathrm{nm}$ are shown in \figurename~\ref{fig2}. 

\begin{figure}
\centering
\includegraphics[width=0.75\columnwidth]{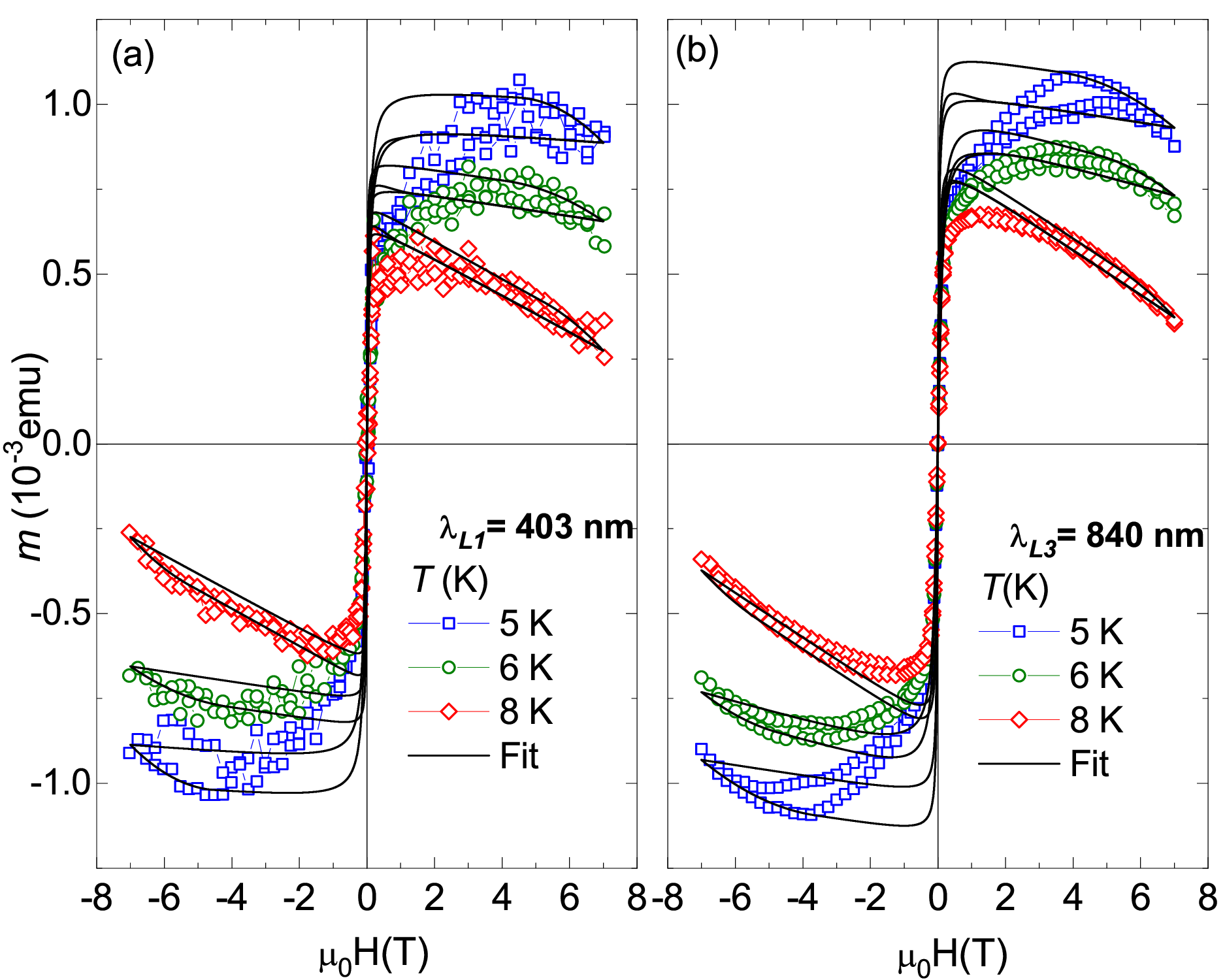}
\caption{\label{fig3} Magnetic moment as a function of the applied field measured at different constant temperatures under illumination. In (a), results measured with $\lambda_{L1}=403$\,nm. In (b), results  measured with $\lambda_{L3}=840$\,nm.  Solid lines are fits done using equation~Eq.~\ref{eq:mtot}.}
\end{figure}

In the dark [\figurename~\ref{fig2}(a,c)], the field dependence of $m(H)$ displays highly unconventional hysteresis behavior that cannot be attributed to typical ferromagnetic or superconducting responses. These findings are consistent with earlier reports~\cite{JBQ_ND-2019,BARZOLA-22}. Under illumination [\figurename~\ref{fig2}(b,d), \ref{fig3}], the hysteresis nearly vanishes, and the magnetic moment exhibits predominantly superparamagnetic characteristics.

\begin{figure}
\centering
\includegraphics[width=0.75\columnwidth]{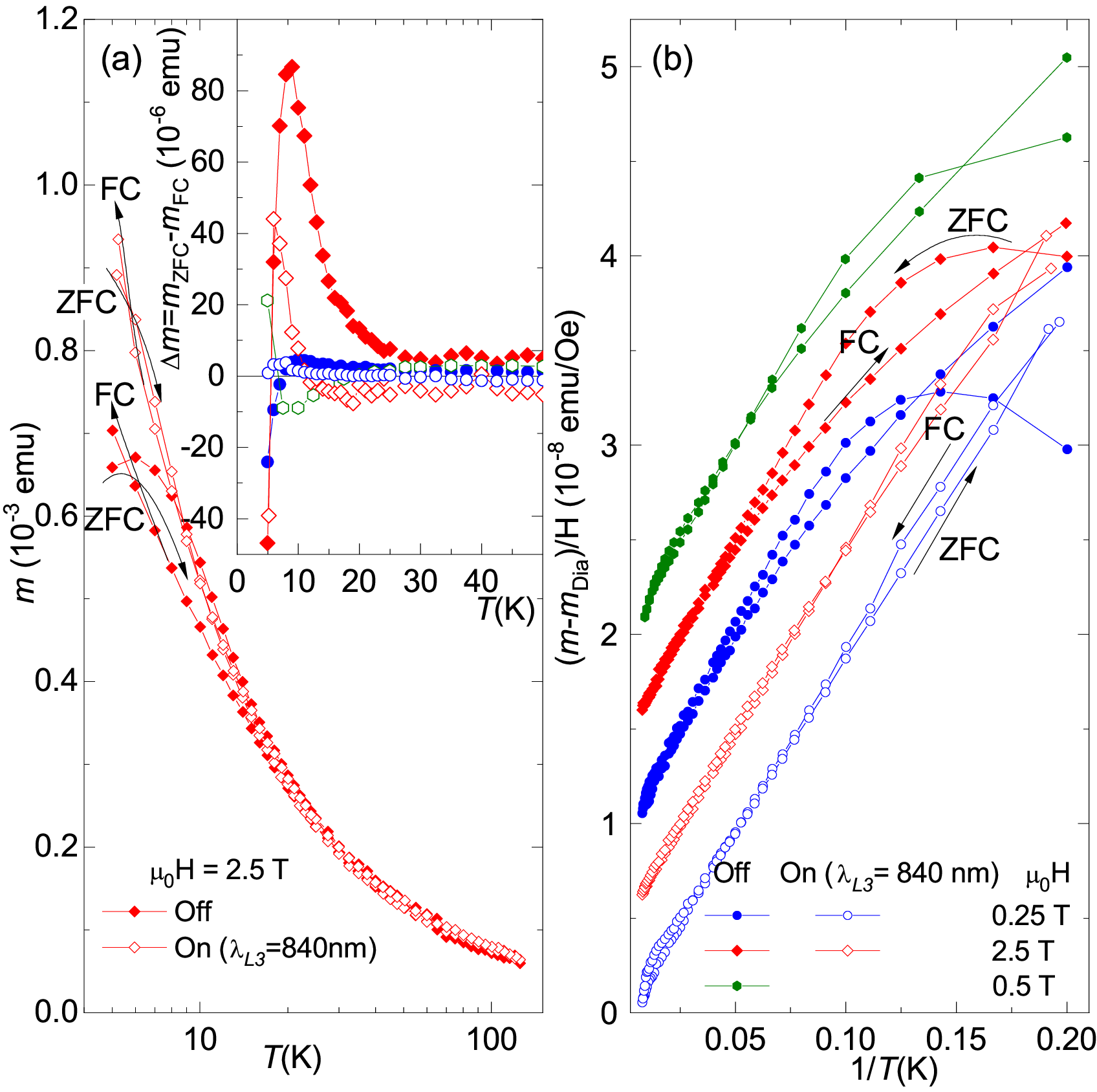}
\caption{\label{fig:fig3a} (a) ZFC and FC results as a function of temperature measured in the dark and during illumination. In the inset the difference $m_{ZFC}(T)-m_{FC}(T)$ is shown. (b) The field susceptibility $\chi(T)$. Results
with full symbols (Off) were obtained in the dark, while results in open symbols (On) were obtained  under illumination ($\lambda_{L3}=840$~nm).}
\end{figure}

The temperature dependence of the magnetic moment, $m(T)$, measured under a constant applied field, is shown in \figurename~\ref{fig:fig3a}. Data were collected both in darkness and under illumination with light of wavelength $\lambda_{L3}=840\,\mathrm{nm}$. In darkness (filled symbols), $m(T)$ follows a predominantly paramagnetic behavior, but below approximately $25\,\mathrm{K}$ a clear irreversibility between the zero-field-cooled (ZFC) and field-cooled (FC) curves emerges. Under illumination (open symbols), this irreversibility almost completely vanishes. 

To emphasize this effect, \figurename~\ref{fig:fig3a}(b) shows the magnetic susceptibility, defined as $\chi(T) = [m(T) - m_{\mathrm{dia}}]/H$, plotted as a function of $1/T$, where $m_{\mathrm{dia}}$ denotes the diamagnetic background measured at $75\,\mathrm{K}$. Upward and downward arrows mark the ZFC and FC data, respectively. For clarity, the curves are vertically shifted by a constant offset. 

Measurements performed under illumination with $\lambda_{L1}=403\,\mathrm{nm}$ and $\lambda_{L2}=633\,\mathrm{nm}$ yielded nearly identical results, whereas those with $\lambda_{L3}=840\,\mathrm{nm}$ showed slightly higher signal intensity and reduced noise. This improvement originates from the longer acquisition time: approximately $440\,\mathrm{min}$ for $\lambda_{L3}=840\,\mathrm{nm}$ compared with $242\,\mathrm{min}$ for the other wavelengths. The time-dependent magnetic moment is analyzed in the later section.

\subsection{Magnetization measurements up to 14\,T}
\begin{figure}
\centering
\includegraphics[width=0.75\columnwidth]{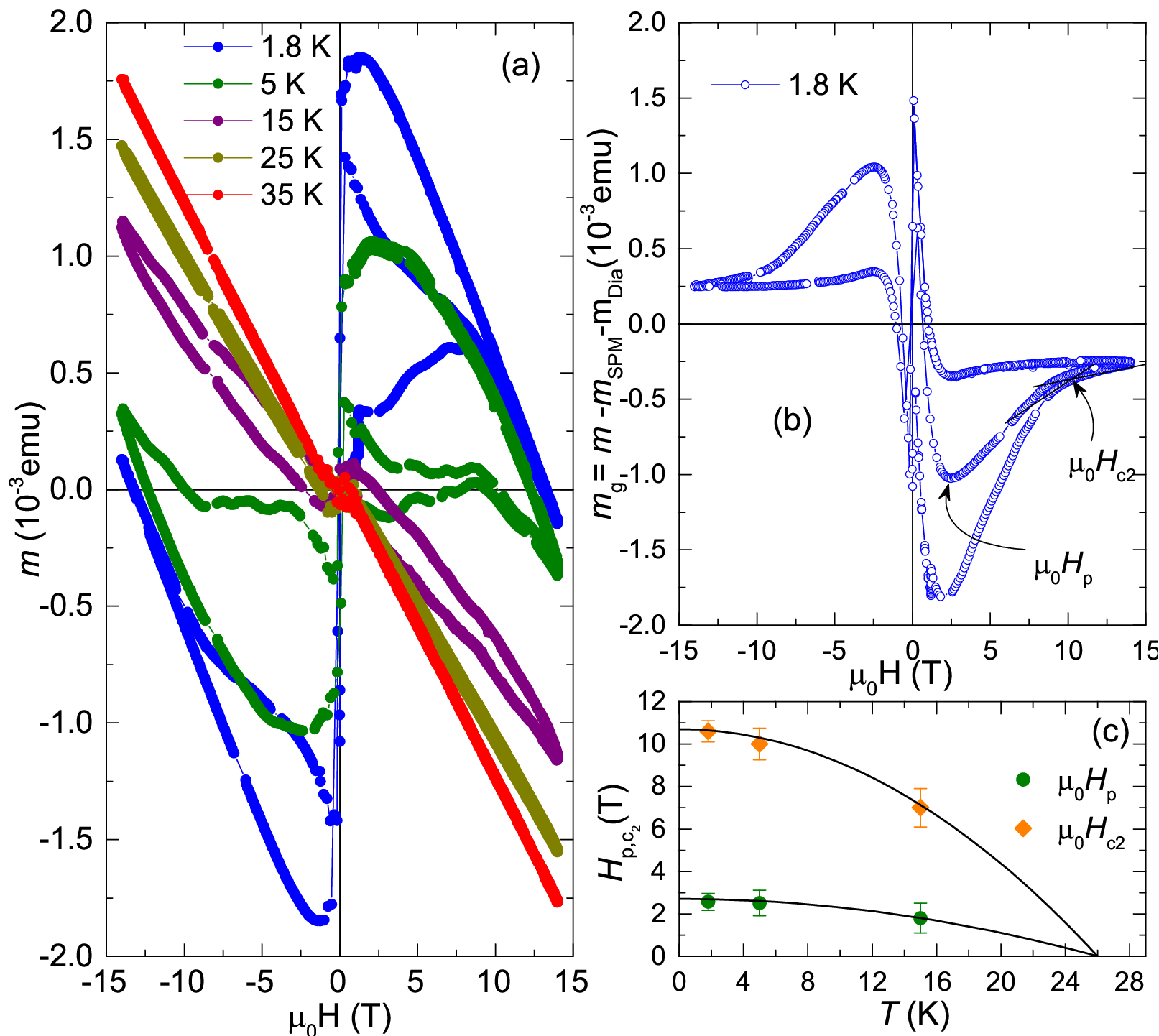}
\caption{\label{fig:figMH14T} Magnetic moment $m$ as a function of the applied field measured at different constant temperatures. (a) As measured, i.e raw data. (b) After removing the superparamagnetic and diamagnetic contributions. (c) Estimated penetration field $\rm{H}_{p}$ and the critical field $\rm{H}_{c2}$.}
\end{figure}

Due to dimensional constraints of the cryostat, which accommodates samples up to approximately $3\times3\,\mathrm{mm^2}$ in surface area, these measurements were performed on a different yellow diamond sample of comparable composition. The magnetization data obtained in darkness at low temperatures are shown in \figurename~\ref{fig:figMH14T}. As seen in \figurename~\ref{fig:figMH14T}(a), the field dependence closely resembles that of the other sample presented in this work [\figurename~\ref{fig2}(a,c)], although the extended field range up to $\pm 14\,\mathrm{T}$ highlights a pronounced diamagnetic contribution. 

The most notable results are, first, that the hysteresis evidently disappears at temperatures above $25\,\mathrm{K}$, consistent with observations reported for similar samples. Second, we observe a marked tendency for the hysteresis to begin closing at fields exceeding $\pm 8\,\mathrm{T}$, consistent with the behavior expected for a Type-II superconductor upon approaching or exceeding the upper critical field $H_{c2}$.

\subsection{Time dependence }

Building on our previous work~\cite{JBQ_ND-2019}, where the unusual magnetic behavior of nitrogen-doped diamond was attributed to coexisting superparamagnetic and superconducting contributions, we investigated the time dependence of the magnetic moment, $m(t)$, under illumination to determine whether light affects either of these components. 

Since the field-dependent $m(H)$ results were nearly identical for the three illumination wavelengths, time-dependent measurements were conducted only for $\lambda_{L2}=633\,\mathrm{nm}$. Before each measurement, the magnetic field was set to zero and the sample was heated to $125\,\mathrm{K}$. After 15 minutes, it was cooled to the target temperature. Once temperature stability better than $10\,\mathrm{mK}$ was achieved, a constant field of $\mu_0H=0.5\,\mathrm{T}$ was applied, and data acquisition began immediately. Each measurement lasted approximately $58\,\mathrm{min}$. After completion, the field was set to zero, the sample was reheated to $125\,\mathrm{K}$, and the sequence was repeated at the next temperature. 

\begin{figure}
\centering
\includegraphics[width=0.75\columnwidth]{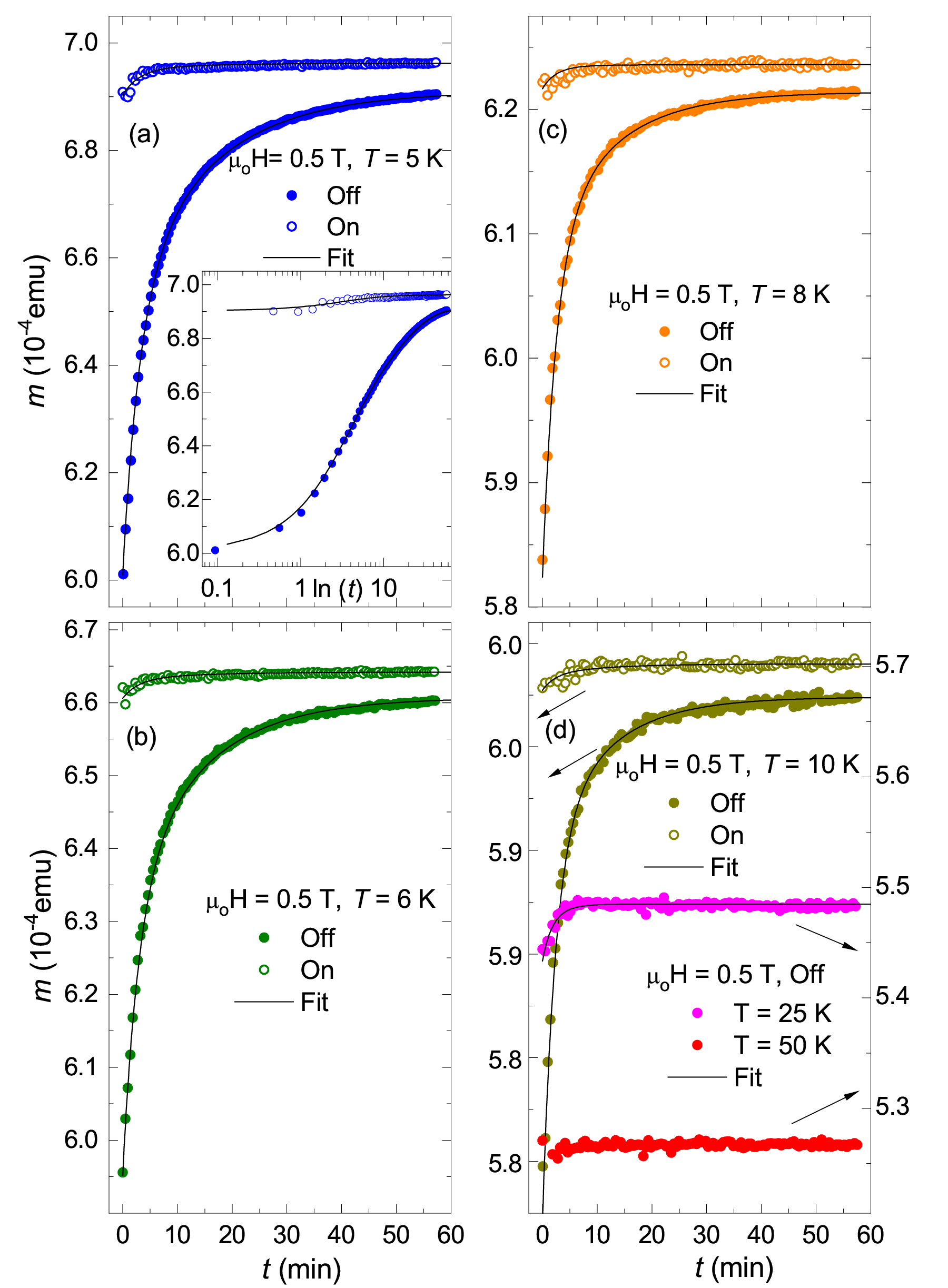}
\caption{\label{fig:fig4} Time dependent magnetic moment measured under an applied field of $\upmu _0 \rm H=0.5$\,T at different constant temperatures. Results under illumination were obtained for $\lambda_L=633$\,nm. The lines
are fits obtained using the equation Eq.~\ref{eq:mtime}.}
\end{figure}

The same procedure was used for measurements in darkness and under illumination, with the only difference being that the laser was switched on at $125\,\mathrm{K}$ and kept on continuously throughout the illuminated runs. The resulting $m(t)$ curves, shown in \figurename~\ref{fig:fig4}, display clear differences between dark (filled symbols) and illuminated (open symbols) conditions. The data obtained in darkness are consistent with those reported in Ref.~\cite{JBQ_ND-2019}, confirming the reproducibility of the results.

\begin{figure}
\centering
\includegraphics[width=0.75\columnwidth]{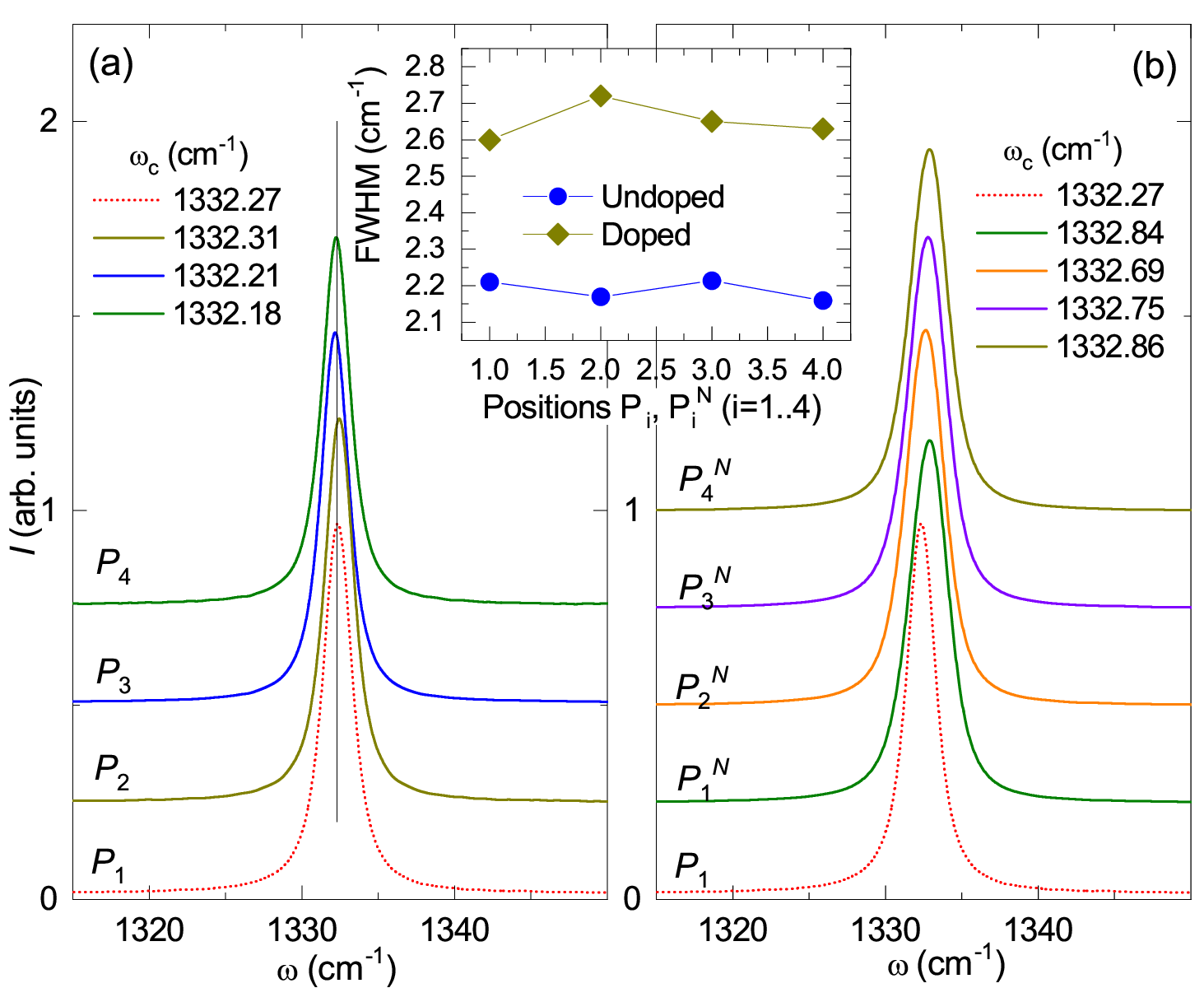}
\caption{\label{fig:fig6}  Raman spectrum measured at room temperature. In (a) results of the undoped sample and in (b) the results of the nitrogen doped sample and one curve from the undoped sample for comparison. The inset is the estimated experimental FWHM of the undoped and doped diamond. The superscript $N$ stands for nitrogen.}
\end{figure}

\subsection{Raman spectroscopy and birefringence microscopy}
We conducted room-temperature Raman spectroscopy at several positions within the sample, approximately 100~µm below the surface, to minimize the influence of surface defects on the spectra, and to assess the uniformity of the vibrational response across the probed region. For reference, we also measured the Raman response of an undoped, single-crystal type~IIa CVD-grown diamond. The results for both samples, recorded at different positions, are shown in \figurename~\ref{fig:fig6}. The center position of the characteristic Raman peak ($\omega_c$) was determined by fitting with a Gaussian function.
Stress birefringence microscopy, a powerful nondestructive method for detecting internal strain in transparent materials, was used to identify strained regions within the diamond. This technique has previously been applied to study strain-induced birefringence in diamond\,\cite{TOLANSKY-66,HOA-14}. The corresponding birefringence images of our sample are presented in \figurename~\ref{fig:fig0}(b--d), while \figurename~\ref{fig:fig0}(a) shows a reference image obtained using a nonpolarized microscope.

\section{Discussion}

Before analyzing the light-induced modifications observed in the $m(H)$, $m(T)$, and $m(t)$ measurements, we emphasize that the observed magnetic changes cannot originate from global heating of the sample.  

This conclusion is supported by three independent arguments. First, all measurements were performed in a cryostat with the sample continuously immersed in liquid $^4$He, providing excellent thermal anchoring. Second, the illumination power was far too low to cause any measurable temperature rise: even under continuous exposure, the absorbed optical power was insufficient to raise the temperature above 25\,K. The sample was irradiated by a uniform, normally incident photon beam (see~\figurename~\ref{fig:fig1}) with an intensity of only $160\,\upmu\mathrm{W/cm^2}$. Under these conditions, and accounting for the thermal conductivity of yellow diamond, the maximum possible temperature increase is estimated to be less than $0.01\,\mathrm{K}$ at temperatures below $25\,\mathrm{K}$ (see the SI).  

Third, the temperature- and time-dependent measurements provide direct evidence against heating. In both cases, the magnetic moment under illumination exceeds that measured in darkness, i.e., $m(T)_\mathrm{On} > m(T)_\mathrm{Off}$ and $m(t)_\mathrm{On} > m(t)_\mathrm{Off}$ (see~\figurename~\ref{fig:fig3a}(a) and \figurename~\ref{fig:fig4}). If illumination merely increased the sample temperature, the opposite trend would be expected, since higher temperature generally reduces the magnetic moment.

As we mentioned above, the field hysteresis measured in the dark is completely unusual compared to a ferromagnetic (FM) or superconducting (SC) hysteresis. To make this clear, we have drawn arrows in \figurename~\ref{fig2}(a), labeled the paths (1, 2, 2$^{\prime}$, 3, 3$^{\prime}$), and divided the figure into quadrants (I, II, III, IV). The path labeled as (1), corresponding to the initial field sweep, resembles a FM or PM response up to the maximum applied field. Path (2) is unusual: first, the magnetic moment increases to a maximum as the applied field decreases; second, it drops abruptly to nearly zero close to zero field. Only the latter behavior is reminiscent of a superparamagnet (SPM).  

When the field is reversed, see path (2$^{\prime}$) in quadrant (III) of \figurename~\ref{fig2}(a), the magnetic moment starts near zero and suddenly increases in the negative direction. This differs from typical FM or SC behavior, where a crossover from quadrant (I) to (III) would pass through quadrant (II). However, the sharp negative increase again resembles an SPM response. Path (3) follows a similar trend as path (2), and the final path (3$^{\prime}$) in quadrant (I) is also unusual, as the magnetic moment retraces path (1) rather than forming a conventional hysteresis loop.  

Illumination has a pronounced effect on the $m(H)$ results shown in \figurename~\ref{fig2}(b,d) and \figurename~\ref{fig3}, leading to a near disappearance of the hysteresis and the emergence of predominantly superparamagnetic behavior. Monochromatic light effectively suppresses the component responsible for the hysteresis, while the remaining small contribution likely originates from regions that retain superconductivity. This observation suggests the presence of at least two distinct superconducting regions in the sample, differing slightly in the size of their clusters and their spatial distribution across the crystal.  

Another notable observation from the $m(H)$ results in \figurename~\ref{fig2}(b,d) and \figurename~\ref{fig3} is that, within experimental accuracy, the magnetic response shows no clear dependence on the wavelength of illumination. In undoped diamond, the band gap is $E_g = 5.45$\,eV, but in nitrogen-doped diamond, the activation energy is difficult to specify due to the variety of nitrogen configurations, including isolated substitutional atoms, aggregates, platelets, and voidites. In particular, substitutional nitrogen in type Ib diamonds is often associated with energy levels approximately 1.7\,eV below the conduction band~\cite{FARRER-69}. Our experiments used photon energies of 1.47, 1.95, and 3.07\,eV, indicating that electronic transitions involving these levels cannot account for the observed changes in $m(H)$.  

These results suggest that the unusual magnetic behavior in yellow diamond is more likely linked to nitrogen aggregates, such as platelets and voidites~\cite{HIRSCH-86}, consistent with theoretical predictions~\cite{MA-05}. Platelets are extended planar defects containing aggregated nitrogen, with sizes ranging from 8\,nm to several micrometers, and nitrogen concentrations within a platelet monolayer spanning 32--46\%~\cite{ALLEN-81,FALLON-95,KIFLAWI-00,SPEICH-18}. Voidites are octahedral, nanometer-scale clusters bounded by $\{111\}$ planes, with diameters between 0.1 and 10\,nm~\cite{HIRSCH-86A}, containing nitrogen in amounts comparable to platelets, typically in molecular N$_2$ form~\cite{KIFLAWI-00,VANTENDELOO-90}. The role of these aggregates in producing the unusual magnetization will be discussed in detail later.

To quantify the illumination-induced changes in the $m(H)$ hysteresis, we refer to the results and interpretation presented in Refs.~\cite{JBQ_ND-2019,BARZOLA-22}, where most of the investigated nitrogen-doped diamond samples were supplied by SUMITOMO~\cite{JBQ_ND-2019} and ELEMENTSIX~\cite{BARZOLA-22}, as in our case. In Ref.~\cite{JBQ_ND-2019}, a simple model was proposed to describe the $m(H)$ behavior of superconducting/ferromagnetic bilayers (LSMO/YBCO and Ni/YBCO) as well as nitrogen-doped diamond. This model successfully reproduced the experimental results~\cite{JBQ_ND-2019,BARZOLA-22}.  

Since the model is based on a linear superposition of diamagnetic, superparamagnetic, and superconducting contributions, it provides a straightforward framework to analyze how these components change under illumination. The model is defined as follows:

\begin{equation}\label{eq:mtot}
m(\text{H})=m_{\text{Dia}}(\text{H})+m_{\text{SPM}}(\text{H})+m_{\text{SC}}(\text{H}),
\end{equation}

where $m_{\rm Dia}(\rm H)$ is the diamagnetic contribution, $m_{\rm SPM}(\rm H)$ the superparamagnetic contribution and $m_{\rm SC}(\rm H)$ is the superconducting contribution. More detailed information about the equations used to fit the $m_{\rm SPM}(\rm H)$ and $m_{\rm SC}(\rm H)$ contributions are given in the main paper and supplementary information in Ref.\,\cite{JBQ_ND-2019}.
The fit results using the equation~Eq.\,\ref{eq:mtot} are shown as continuous lines in \figurename\ref{fig2} and \figurename\ref{fig3}. From the fit results we can observe that the model describes well the experimental results of the measured $m(\rm H)$ in the dark and under illumination.  
\begin{figure}
\centering
\includegraphics[width=0.75\columnwidth]{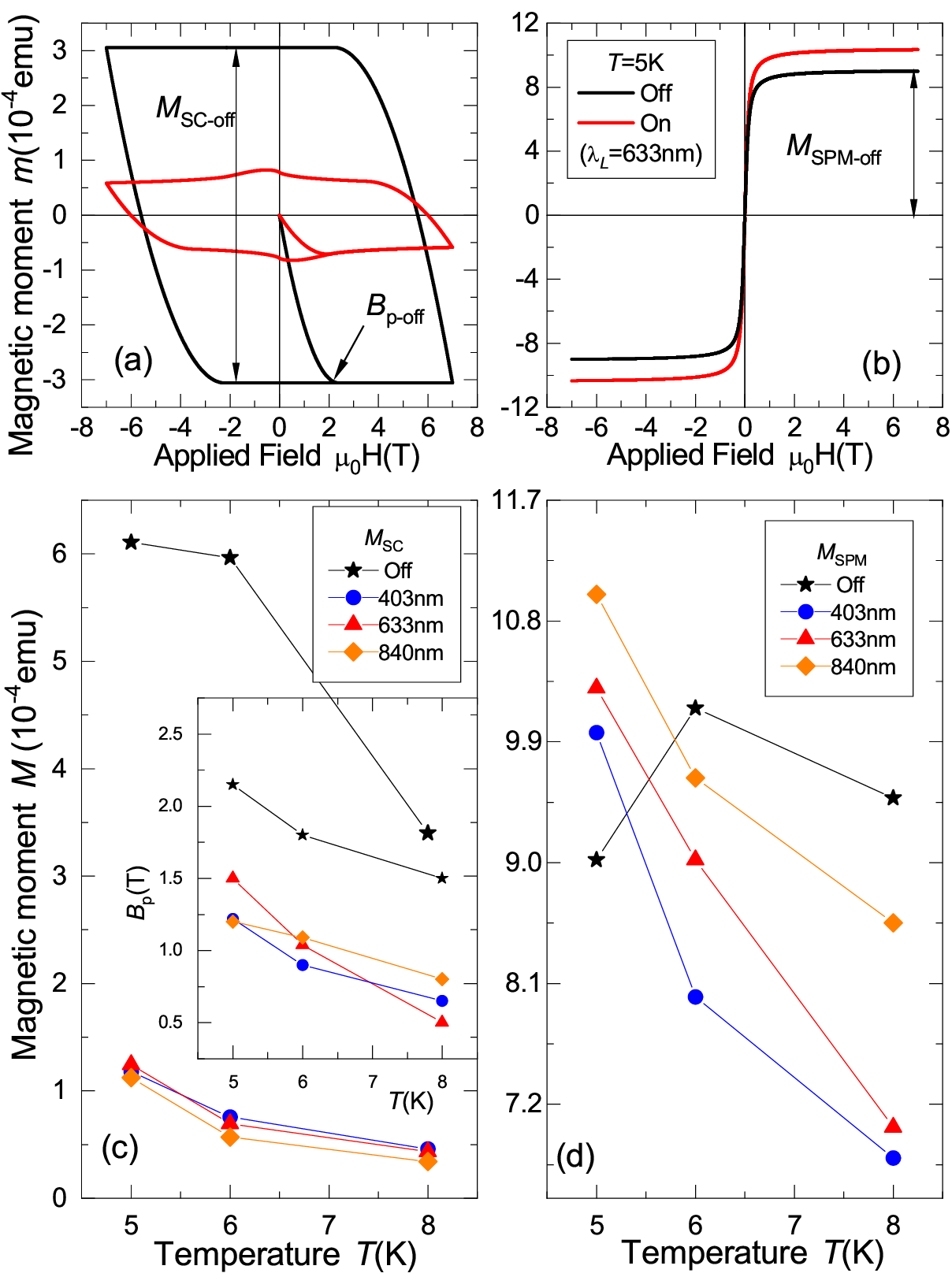}
\caption{\label{fig:fig9} Parameters obtained from the fits to the $m(\rm H)$ results measured without and under illumination. In (a,b) hysteresis loop models  for the SC and SPM contributions are shown . In (c,d) the parameters $m_{\rm SC}$ and $m_{\rm SPM}$ vs. temperature respectively are plotted. The inset in (c) is showing the penetration field $B_p$ fit parameter. The size of the symbols corresponds to the error bars that allow for a good fit.}
\end{figure}
Changes with and without illumination of the superconducting $m_{\rm SC}\rm (H)$ and superparamagetic $m_{\rm SPM}\rm (H)$ contributions at $T=5$\,K are plotted in\,\figurename\ref{fig:fig9}\,(a,b).
A summary of the relevant obtained fit parameters are also plotted in\,\figurename\ref{fig:fig9}\,(c,d). From the experimental results and the fits, we can conclude that the illumination of the sample has substantially changed the contribution responsible for the superconducting hysteresis, according to our interpretation. The quotient of the parameters (see \figurename\ref{fig:fig9} ) are $M_{\rm SC-Off}/M_{\rm SC-On} \approx 6$, $B_{\rm p-Off}/B_{\rm p-On} \approx 1.5$ and $M_{\rm SPM-Off}/M_{\rm SPM-On}\approx 1.1$, indicating that the more affected contribution is the $m_{\rm SC}$.
The change in the superparamagnetic contribution is small but real, this statement is more evident in the time dependent results which we discuss later. 
%considering the differences during measurements, e.g. time dependence during measurements and the used power are a bit different, we can %conclude that the superparamagnetic contribution remains almost unchanged after illumination.  

Another unusual feature appears in the temperature dependence of the magnetic moment, $m(T)$. As shown in \figurename~\ref{fig:fig3a}(a), the results reveal a clear difference between measurements taken in darkness and under illumination. The irreversibility observed in the dark almost disappears when the sample is illuminated. This effect becomes more evident from the difference $\Delta m(T) = m_{\mathrm{ZFC}}(T) - m_{\mathrm{FC}}(T)$, calculated for both conditions. The inset of \figurename~\ref{fig:fig3a}(a) displays these differences, where full and open symbols correspond to measurements in darkness and under illumination, respectively.  

In both cases, $\Delta m(T) \approx 0$ for $T \gtrsim 25$\,K. Under illumination, however, $\Delta m(T)$ vanishes already below $T = 10$\,K at the highest applied field and remains nearly zero across all temperatures at the lowest field. The small residual $\Delta m(T)$ under illumination likely originates from localized regions that still exhibit hysteresis, consistent with the $m(H)$ data at $\upmu_0 H = 2.5$\,T shown in \figurename~\ref{fig3}.  

Previous studies of nitrogen-doped diamond at low temperatures~\cite{JBQ_ND-2019,BARZOLA-22} identified type-II superconducting behavior, but the limited field range ($\pm7$\,T) precluded direct observation of the upper critical field $H_{c2}$. Figure~\ref{fig:figMH14T}(a) presents $m(H)$ measurements extended to $\pm 14$\,T. As discussed earlier, the data for $\upmu_0H > 8$\,T indicate that the sample begins to approach $\upmu_0 H_{c2}$. To illustrate this more clearly, Figure~\ref{fig:figMH14T}(b) shows, as an example, the $m(H)$ curve at $T=1.8\,\mathrm{K}$ after subtracting the diamagnetic and superparamagnetic contributions. The procedure used for this subtraction, together with additional details, is described in the SI document. In a granular Type-II superconductor \cite{CLEM-88}, both $\mu_0H_p$ (the full penetration field) and $\mu_0H_{c2}$ (the upper critical field) are fundamental characteristic fields. The penetration field $\mu_0H_p$ refers here to the intergranular full penetration field and is defined as the external magnetic field at which magnetic flux fully penetrates the network of weak links between the superconducting grains of the sample. In contrast, $\mu_0H_{c2}$ is the intrinsic magnetic field at which superconductivity is completely suppressed within the individual grains.

Figure~\ref{fig:figMH14T}(b) clearly shows these two characteristic fields, $\mu_0H_p$ and $\mu_0H_{c2}$, indicated by arrows. While $\mu_0H_p$ can be determined relatively easily, $\mu_0H_{c2}$ is obtained from the intersection of the lines indicated in the figure. Determining the lower critical field $\mu_0H_{c1}$ in our samples is more difficult because the $m_{\mathrm{SPM}}$ contribution at these field strengths introduces considerable uncertainty into its determination. Nevertheless, based on Figure~\ref{fig:figMH14T}(c), we can conclude that $\mu_0H_{c1}<0.5\,\mathrm{T}$. The resulting temperature dependences of $\mu_0H_p(T)$ and $\mu_0H_{c2}(T)$ are plotted in Figure~\ref{fig:figMH14T}(c). These temperature dependences can be described empirically by the relation
\[
H_{p,c}(T)=H_{p,c}(0)\left[1-\left(\frac{T}{T_c}\right)^2\right],
\]
where $\mu_0H_{p,c}(T)$ denotes the characteristic field at a given temperature $T$, $\mu_0H_{p,c}(0)$ is the corresponding characteristic field extrapolated to zero temperature, and $T_c$ is the critical temperature. The corresponding fits are shown as solid lines in Figure~\ref{fig:figMH14T}(c), yielding $\mu_0H_p(0)=2.7\,\mathrm{T}$ and $\mu_0H_{c2}(0)=10.7\,\mathrm{T}$, with a common critical temperature of $T_c=26\pm2\,\mathrm{K}$. The obtained $T_c$ value agrees well with the values derived from the measurements presented in this work, as well as with our previous results \cite{JBQ_ND-2019,BARZOLA-22} and those reported by other authors \cite{BUGA-24}.

The estimated upper critical field $\mu_0H_{c2}$ is comparable to values reported for boron-doped diamond superconductors \cite{TAKANO-07,OKAZAKI-15}. The results shown in Figure~\ref{fig:figMH14T}(b) also allow us to estimate several intrinsic properties of the individual grains. At $T=1.8\,\mathrm{K}$, we obtain $\mu_0H_{c2}\approx10.6\,\mathrm{T}$ and $\mu_0H_p\approx2.6\,\mathrm{T}$, a value very similar to that obtained from the fit performed on the large sample, as can be seen in the inset of Fig.~\ref{fig:fig9}(c). Using the upper critical field $\mu_0H_{c2}$, the Ginzburg--Landau coherence length $\xi_{\mathrm{GL}}$, which represents a fundamental length scale associated with the superconducting condensate and vortex cores \cite{GINZBURG-55}, can be determined from the standard relation
\[
\xi_{\mathrm{GL}}=\left(\frac{\Phi_0}{2\pi\mu_0H_{c2}}\right)^{1/2},
\]
where $\Phi_0$ is the magnetic flux quantum. We obtain $\xi_{\mathrm{GL}}\approx5.6\,\mathrm{nm}$, which is comparable to values reported for superconducting boron-doped diamond \cite{EKIMOV-04,TAKANO-07}.

Using the $m_g(H)$ data and the value of $\mu_0H_p$ obtained at $1.8\,\mathrm{K}$, we can further estimate the grain radius $R_g$ using the Bean model for an isolated spherical grain. The grain radius can be estimated from the full penetration field according to
\[
\mu_0H_p=\mu_0J_cR_g,
\]
\noindent where $V_{\mathrm{total}}$ is the total sample volume and $J_c$ is the critical current density \cite{PEREZ-RODRIGUEZ-01}. However, applying this relation requires an estimate of $J_c$. For this purpose, we use a value reported in the literature. In samples similar to ours, investigated by Buga et al. \cite{BUGA-24}, $J_c$ was estimated to be approximately $2\times10^{10}\,\mathrm{A/m^2}$ at low temperatures. Using this value, we estimate
\[
R_g=\frac{2.5\,\mathrm{T}}
{4\pi\times10^{-7}\,\mathrm{H/m}\times2\times10^{10}\,\mathrm{A/m^2}}
\approx99\,\mathrm{nm},
\]
corresponding to a grain diameter of approximately $198\,\mathrm{nm}$.

Based on the obtained $R_g$ and the hysteresis data shown in Figure~\ref{fig:figMH14T}(b), we can also estimate the critical current density of our sample using
\[
\Delta m_g=\frac{2}{3}J_cR_g,
\]
\noindent where $V_{\mathrm{total}}$ is the total sample volume and $\Delta m_g$ is estimated from the difference between the lower and upper branches of $m_g(H)$, evaluated at $\mu_0H=2.5\,\mathrm{T}$, where the two branches run approximately parallel. Further details of this calculation are provided in the SI document. We obtain $J_c=5.84\times10^9\,\mathrm{A/m^2}$, which is approximately one order of magnitude lower than the value estimated by Buga et al. \cite{BUGA-24}.

Overall, the hysteresis behaviour of $m_g(H)$ and the superconducting parameters derived from the $m_g(H)$ measurements are comparable to those reported for boron-doped diamond films. For example, Kumar et al. reported values of $\mu_0H_{c2}\approx11.9\,\mathrm{T}$, $\xi_{\mathrm{GL}}=7.2\,\mathrm{nm}$, and $J_c\approx10^{11}\,\mathrm{A/m^2}$ \cite{KUMAR-18}. The consistency of these characteristic parameters with previously reported values further supports our interpretation of the observed behavior in terms of granular superconductivity in our samples.

% and/or also perhaps from the superparamagnetic regions themself, because irreversibility is also observed in pure superparamagnetic %samples~\cite{PAULA-16}. 
As we mentioned before and as it is shown in Ref.\,\cite{JBQ_ND-2019}, the low N doped diamond samples show a time dependence in the magnetic moment, indicating the existence of flux creep, similar as in type II superconductors~\cite{ANDERSON-62,ANDERSON-64,GUREVICH-93,HOEKSTRA-98}. 
The time dependent $m(t)$ results in the dark are observable 
 up to a maximal temperature, which in our sample disappears at temperatures $T\gtrsim 25$\,K and evidently at $T=50$\,K, as it is shown in the results of \figurename\ref{fig:fig4}\,(d).  
 From our results shown in \figurename\ref{fig:fig4} it is clear to observe a time dependent $m(t)$ in the dark, but it changes considerably when the sample is illuminated. In order to get information about the changes with and without illumination we have fitted the $m(t)$ results using the following equation:  
\begin{equation}\label{eq:mtime}
m(t)=m_{0}+m_{1}\exp(-t/\tau_1)+m_{2}\exp(-t/\tau_2),
\end{equation}

\begin{figure}
\centering
\includegraphics[width=0.75\columnwidth]{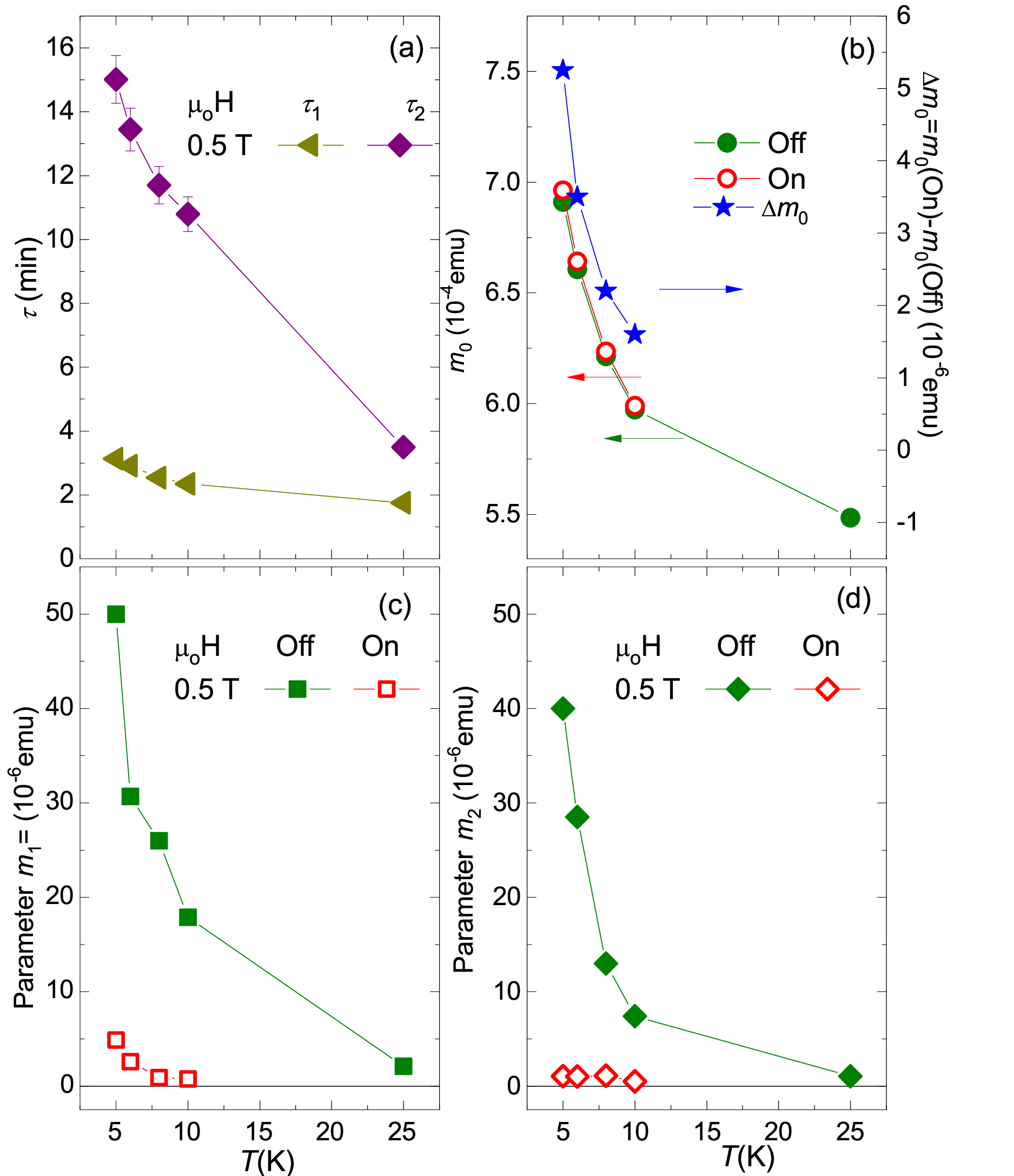}
\caption{\label{fig:fig8}Parameters obtained after fitting the time dependent magnetic moment $m(t)$ in the dark and after illumination with light of $\lambda_L=633$\,nm. The size of the symbols corresponds to the error bars that allow for a good fit.}
\end{figure}
with $m_0$ being the magnetic moment when the time $t\rightarrow \infty$, $m_1$ and $m_2$ prefactors and $\tau_1$ and $\tau_2$ relaxation times.
We employed two distinct relaxation times in the fitting equation, which is reasonable given that the magnetic moment evolving under illumination likely originates from at least two main contributions at temperatures $T \lessapprox 25$\,K. The solid lines in \figurename~\ref{fig:fig4}, and more clearly in the inset of \figurename~\ref{fig:fig4}(a), demonstrate that Eq.~\ref{eq:mtime} reproduces the experimental data remarkably well.  

During the fitting process, both with and without illumination, we assumed that the relaxation times $\tau_1$ and $\tau_2$ remain constant. Allowing them to vary freely under illumination yielded nearly identical values to those obtained in darkness, confirming that illumination does not significantly affect the relaxation dynamics. Therefore, the light-induced effects can be analyzed through the changes in the parameters $m_0$, $m_1$, and $m_2$, which are plotted as functions of temperature in \figurename~\ref{fig:fig8}.  

As shown in \figurename~\ref{fig:fig8}(a), the obtained $\tau_1$ values are much larger than $\tau_2$ for the applied field used, and both decrease almost linearly with increasing temperature. A similar temperature dependence of the relaxation time has been observed in high-$T_c$ superconductors such as YBa$_2$Cu$_3$O$_7$~\cite{GUREVICH-93}, exhibiting a near-linear trend followed by a sharp change close to the critical temperature $T_c$.  

The parameter $m_0$, presented in \figurename~\ref{fig:fig8}(b), is nearly independent of illumination, though slightly higher values appear under light exposure. This difference, quantified as $\Delta m_0 = m_{0,\mathrm{On}} - m_{0,\mathrm{Off}}$ and shown in the inset of \figurename~\ref{fig:fig8}(b), is positive ($\Delta m_0 > 0$). Such an increase can be attributed to regions that were superconducting in darkness and become paramagnetic upon illumination—a characteristic response of superconductors above $T_c$. Considering the trends observed in $m(H)$ and $m(T)$, we associate $m_0$ primarily with the superparamagnetic contribution.  

In darkness, the parameter $m_1$ is clearly larger than $m_2$, although both exhibit similar temperature dependence. Under illumination, both parameters are strongly reduced: $m_1$ shows only weak temperature dependence, while $m_2$ becomes nearly temperature independent, as shown in \figurename~\ref{fig:fig8}(c,d). These behaviors indicate that $m_1$ and $m_2$ represent the magnetic response of two distinct superconducting regions. Their differing magnitudes in darkness and systematic reduction under illumination provide compelling evidence for the coexistence of two dominant superconducting regions within the sample.

The influence of weak visible light on flux creep in high-temperature superconductors such as YBCO\,\cite{LANDAU-95} and La\textsubscript{1.8}Ba\textsubscript{0.2}CuO\textsubscript{4}\,\cite{LAIHO-90} has been investigated at low temperatures. In YBCO films, visible light with intensities below 1000\,$\upmu$W/cm$^2$ was found to drastically accelerate flux creep at liquid-helium temperatures\,\cite{LANDAU-95}. Similarly, Yurgens and Zavaritsky\,\cite{YURGENS-92} reported that visible light with an intensity of 100\,$\upmu$W/cm$^2$ significantly affected magnetic relaxation in the high-temperature superconductor Bi-2215.

We can state that the $m(t)$ results without and under illumination are a clear proof of the existence of flux creep in our sample, similar as observed in typical type II superconductors. 

In early studies on the destruction of superconductivity by laser light of energy $E_L\gg 2\Delta$, with $\Delta$ the superconducting gap, e.g. done on thin films of Pb at temperatures below the usual $T_c$, it was concluded that the destruction is not due to normal lattice heating effect\,\cite{TESTARDI-71}. The explanation is, the electron gas may be heated above the lattice temperature by the light absorption, this overheating of the electron gas is due to the breaking of Cooper pairs under the illumination, so the extra unpaired electrons depress the energy gap
and raise the free energy of the superconducting state with a disappearance of superconductivity at
temperatures below the ordinary $T_c$~\cite{OWEN-72,TESTARDI-71,VUJICIC-75,YURGENS-92,PARKER-75}.

For that reason, due to the similarity of the results collected in our experiment with the results observed in type II high temperature superconductors, we believe
that the changes in the magnetic moment under illumination in our experiments are mainly because of the superconducting properties  
 present in some regions dispersed within the sample. 
%The changes of the flux creep parameters $m_1$ and $m_2$ under illumination, we can attribute to the inelastical scattering of  
%phonons produced during the recombination of the unpaired electrons with the vortices, leading to a weakening of the pinning forces. 

\begin{figure}
\centering
\includegraphics[width=0.65\columnwidth]{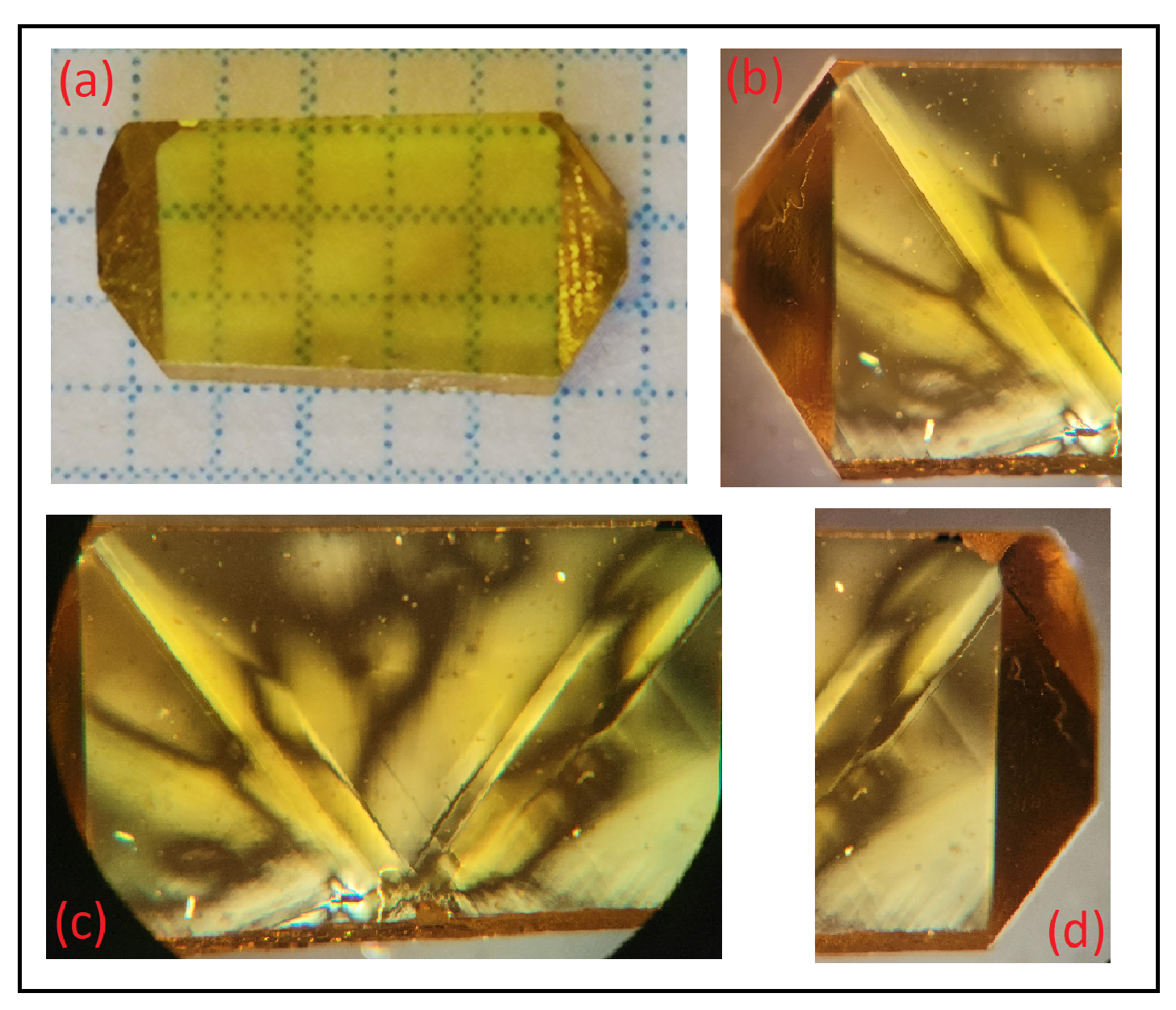}
\caption{\label{fig:fig0} (a) Optical image of the main surface of the sample. (b)-(d) Optical birefringence images of the same sample, showing the left, middle, and right region, respectively.}
\end{figure}

%A open question remains where the superconductivity and the superparamagnetism are coming from. 
Superconductivity in diamond  as consequence  of the doping with nitrogen was already theoretically studied by Ma~\textit{et al.}\,\cite{MA-05}. From these calculations they have predicted for a doping concentration of
$5\,\%$ ($\approx 0.905 \times 10^{28}$m$^{-3}$) a critical temperature of $T_c=23.2$\,K can be reached, which is signiﬁcantly higher for the similar doping concentration when  diamond is p-doped ( $T_c=12.3$\,K), for example with Boron. Recent theoretical
 studies done by Costa~\textit{et al.}\,\cite{COSTA-21} support previous calculations\,\cite{MA-05}, additionally 
 they have  reported that when diamond with a doping of $15\,\%$ ($\approx 2.7 \times 10^{28}$m$^{-3}$) nitrogen is exposed to external pressure ($P\geq 10$\,GPa) a significant increase of the $T_c>110$\,K is predicted \,\cite{COSTA-21a}.  
Considering that the global nitrogen  content in our sample is $\approx 1-7.1\times10^{25}$m$^{-3}$, a doping concentration many orders smaller
 than used in the theoretical calculations, we observe clear signs of superconductivity but in coexistence with superparamagnetism. To remark, the theoretical calculation assumes a homogeneous distribution of nitrogen, but our sample  is highly inhomogeneous regarding the nitrogen doping, as evident by the color variation observed in the optical picture shown in \figurename\ref{fig:fig0}(a). 
Therefore, we expect that in some regions of our sample 
 an agglomeration (e.g as platelets and/or voidites) of nitrogen can be found, which increases the local doping concentration. Another point to consider is the role of pressure on the $T_c$, and as the theoretical works predict, it can be considerably increased. 
Raman spectroscopy was used to characterize the influence of nitrogen content in doped samples by analyzing the changes in the Raman characteristic peak position and changes in the full-width at the half maximum\,\cite{PRAWER-04}. In the literature it was observed that
diamond samples containing 500\,ppm nitrogen, have a negligible shift of the characteristic peak, i.e. no measurable effects of the nitrogen
concentration on the Raman line position or anharmonicity were found\,\cite{SUROVTSEV-17}.  
Hanzawa et-al\,\cite{HANZAWA-96}, investigated the change of the FWHM of synthetic Ib diamonds and on basis
of their experimental results,  an empirical formula to correlate the FWHM and the nitrogen 
content in ppm was proposed, and it reads as following:
\begin{equation}\label{eq:nppm}
\rm FWHM(1/cm)=2.14+1.23\times10^{-3}\,N(ppm)
\end{equation}
\\
Thus, from our experimental results shown as inset in  \figurename\ref{fig:fig6}, we can assume as average $\rm FWHM = 2.17\,cm^{-1}$ and  $\rm FWHM = 2.65\,cm^{-1}$ for the undoped and doped diamond, respectively. Applying the equation Eq.\,\ref{eq:nppm} we estimate that our investigated samples have a nitrogen content of  8\,ppm for the undoped and 415\,ppm for the doped samples. The N content estimated using Raman results of the undoped CVD growth diamond is in the order of the content according to the company information, where a nominal value of content of $<\,5$\,ppm was given. On the other hand, in the case of the doped sample, the value obtained is higher than the nominal value given by the company, although similar difference with respect to the nominal value was reported in\,\cite{JBQ_ND-2019}, where the N content was measured in similar samples as ours, by means of infrared spectroscopy. Doped\,\cite{PRAWER-04} and undoped\,\cite{HANFLAND-86} diamond samples
can show a shift to a higher wave number (blue shift) when the sample contains and or is under compressive stress. 
Taking an average of many results in the literature \cite{GRIMSDITCH-78,BOPPART-85,AGER-93}, we assume that the rate of the
 Raman line shift is about $\rm 2.5\,cm^{-1}/GPa$, and assuming from our results the average shift 
$\Delta \omega \rm = 0.785\,cm^{-1}$, we obtain a persistent stress in our sample of
$P=0.314$\.GPa, a value very similar to the one reported in the literature\,\cite{JASBEER-16}.
To resume, in our Raman results we have observed an increase in the FWHW and an evident blue shift of the characteristic Raman peak, indicating that our sample has a mean value of 415\,ppm nitrogen dispersed everywhere and also a mean value of internal pressure of 0.314\,GPa, 
which probably was originated during the  HPHT method growth process.  

It is well known that birefringence microscopy is a powerful method for detecting internally strained regions in diamond~\cite{LANG-67,ENCKEVORT-88} and other materials.
It was experimentally shown that the presence of nitrogen in diamond causes birefringence, because it modifies the lattice parameter. The mismatch of the lattice parameters between growth regions with different nitrogen concentrations results in birefringence being generated at these regions\,\cite{HOWELL-12}(and literature therein). 
Birefringence studies done on N-doped diamond have concluded that regions with the smaller nitrogen concentration are under tension, while the regions with the larger nitrogen concentration are under compression. 
We believe it is worthy to mention that using strain birefringence, it was possible to estimate the remnant pressure on an inclusion (e.g. calcite, omphacite and coesite) in diamond, thus the remaining internal pressure was estimated to be between 0.34 to 3.6 GPa~\cite{BARRON-08,HOWELL-12a}, which are very similar values as we estimated in our sample using the Raman results. If such amount of pressure exists in our sample, it can be important for the superconductivity, unfortunately in the literature experimental results in similar samples as used in this work, still don't exist.
Considering that inhomogeneous nitrogen concentration induces  strain in the diamond lattice, and the combination of high pressure with nitrogen doping induce/increase superconductivity in diamond, we have investigated the existence of stress in our sample.
Birefringence images are shown in \figurename\ref{fig:fig0}\,(b-d) in comparison with a simple optical image  in \figurename\ref{fig:fig0}\,(a). The results show evident birefringence patterns in our sample and having in mind that in a cross polarized image, the dark regions means optical isotropy (because its strain free) and the illuminated regions strain, we can conclude that our
sample presents mainly internal strain. 
 Thus, a combination of locally increased nitrogen concentration and the local stress can be responsible for the superconductivity   observed in this work. For the superparamagetic contribution we don't have a clear explanation, and no theoretical prediction about its origin in correlation with nitrogen doping exists. 
We find it important to note that Remes\,\textit{et al.}\cite{REMES-15} reported the observation of ferromagnetism in nitrogen-implanted diamond films at implantation doses of $7\times 10^{11}\rm ions/cm^2$ and  $7\times 10^{14}\rm ions/cm^2$. Their conclusions were based on magnetic circular dichroism (MCD) and magnetization measurements using a superconducting quantum interference device (SQUID). They attributed the observed ferromagnetism to structural distortions, proposing that the crystal field induced by nitrogen-related dipole moments lifts the degeneracy of energy levels, thereby giving rise to magnetic ordering. However, we find it important to point out that their magnetization data at $T=5$\,K (Figure 3 in their paper) appear to exhibit characteristics more consistent with paramagnetism rather than ferromagnetism.

Example of coexistence of ferromagnetism and superconductivity was reported in nanoparticles of Sn~\cite{LI-08}, where a ferromagnetic spin polarization near the particle surface coexists with superconductivity inside the particle. In nanoparticles of YBCO~\cite{HASANAIN-11,ZHU-12}, the ferromagnetism is attributed to the presence of surface oxygen vacancies. The coexistence of superconductivity and ferromagnetism in materials containing $d$- and $f$-electrons has been known for several decades, with well-established examples such as ZrZn$_2$\,\cite{PFLEIDERER-01}, UGe$_2$\,\cite{SAXENA-00}, Sr$_{0.5}$Ce$_{0.5}$FBiS$_2$\,\cite{LI-15}, EuFe$_2$As$_2$\,\cite{JIN-19}, and DyMo$_6$S$_8$\,\cite{MONCTON-78}. More recently, this coexistence has also been discussed for carbon-based materials. In particular, ABC-stacked multilayer graphene has been predicted theoretically\,\cite{AWOGA-23} and observed experimentally\,\cite{HAN-25} to host both superconducting and ferromagnetic states. Similar behavior has been reported in bundles of double-wall carbon nanotubes\,\cite{JBQ-15}. In contrast, for compounds composed solely of carbon and nitrogen, such coexistence has so far only been predicted theoretically for a monolayer of graphitic carbon nitride (C$_3$N$_4$)\,\cite{BRANGI-25}.

Considering previous results\,\cite{JBQ_ND-2019,BARZOLA-22} obtained from more than ten different samples (all measured in darkness), together with the findings presented here, we propose the following picture for the formation of SPM/SC regions in N-doped diamond: during crystal growth, small regions with locally high nitrogen concentration and significant lattice stress develop within the crystal.
 These can be thought of as small and semispherical regions dispersed in the sample, because of the lack of anisotropy observed in the magnetic moment,  
but with the characteristic that in the core of such regions there is a high concentration of nitrogen and less at the surface. So when the sample temperature is less than $T\lessapprox 25$\,K in the dark, because of the new physical properties imposed by the high nitrogen content and local stress,  the core part goes into the SC state and at the same time the surface part into a SPM state.
This suggests an intimate physical mechanism that occurs simultaneously, explaining why both states coexist within the same temperature range. A similar behavior has been observed in ZrZn$_2$\,\cite{PFLEIDERER-01}, where the ferromagnetic state appears to be a prerequisite for superconductivity, as both phases vanish at the same critical pressure.
 During illumination the SC contribution is suppressed while the SPM survives, because even though the sample is illuminated (Cooper pairs are broken), the physical  process which creates the SC and SPM still is present.
To emphasize, the SPM is not a contribution coming from the ferromagnetic contaminants as Fe, Ni and Co, which are in the order of tenths of ppm, as it was already investigated experimentally in Ref.\,\cite{JBQ_ND-2019}.    
We believe that the changes obtained after illumination support our interpretation of the coexistence of 
superparamagnetism and superconductivity in N-doped diamond, which are very similar as those observed in well known superconductors such as Pb\,\cite{TESTARDI-71}, LaBaCuO\,\cite{LAIHO-90} and YBaCuO\,\cite{LANDAU-95,LAIHO-94}.

In this work we are for the first time presenting experimental results showing changes  of the magnetic properties under illumination of a commercial  diamond sample doped with nitrogen.
Considering the  previous results, and the new results presented in this work,  we can say with great certainty that N-doped diamond 
with nitrogen content between 85~ppm and 415~ppm, which is the maximum doping level studied to date, shows the coexistence of superparamagnetism and  superconductivity.

\section{Conclusions}
The magnetic properties of a N-doped diamond were investigated in the dark  and under illumination with monochromatic light of different wavelengths. 
In the dark, the investigated sample shows unconventional magnetic moment behavior in field, temperature and time dependence with
 a critical temperature of $T \lesssim 25$~K and above it, the sample behaves as a typical non-doped diamond.
Evidences of mutual coexistence of superparamagnetism and superconductivity observed in the dark are in agreement with previous 
results performed in the dark. Upon illumination, regardless of the wavelength used in this study, the superconducting contribution nearly vanishes, whereas the superparamagnetic component remains largely unaffected or shows a slight increase. Considering the fact that we used a very small intensity of light, 
we discard the lattice heating as origin of the changes observed in our results.
 The changes of the magnetic properties
  during illumination we attribute to the breaking of Cooper pairs which reduce the critical temperature and weaken the pinning forces by inelastic scattering of the unpaired electrons with the vortices.  
 This leads us to conclude that the results under illumination support the interpretation that the unusual magnetic behavior observed in low doped N-diamond are  because of the coexistence of superparamagnetism and superconductivity existing in some regions of the sample.  
 By using a simple model, which considers the superposition of a diamagnetic, superparamagnetic  and superconducting contribution, we can support our conclusions. 
The origin of superconducting and superparamagnetic regions in our sample we attribute to those regions containing high amount of
nitrogen in comparison to the rest of the sample and the additional presence of internal stress,
 verified by birefringence measurements. Finally, magnetization  results up to 14\,T show clear indications that the sample has reached the second critical field and the values obtained are very similar to those of the superconducting boron doped diamonds.   
All experimental results presented in this work were obtained using commercial low nitrogen doped diamond, 
 not yet reported in the literature. Thus far, only limited theoretical predictions or explanations have been proposed. We hope this work will motivate further experimental and theoretical efforts to elucidate the underlying mechanisms and explore possible future applications.

\section*{Data availability statement}
The datasets generated during and/or analysed during the current study are available from the corresponding author on reasonable request.
\section*{Competing interests}
The authors declare no competing interests.
\section*{Funding}
J.B-Q. was supported by the Collaborative Research Center SFB 762 ‘Functionality of Oxide Interfaces’. We acknowledge the support from the W\"{u}rzburg-Dresden Cluster of Excellence on Complexity and Topology in Quantum Matter--$ctd.qmat$ (EXC 2147, Project No.\ 390858490).

\acknowledgements We acknowledge the support of the HLD at HZDR, member of the European Magnetic Field Laboratory (EMFL).

\bibliography{nd}

\end{document}